\documentclass{ecai} 

\usepackage{latexsym}
\usepackage{amssymb}
\usepackage{amsmath}
\usepackage{amsthm}
\usepackage{booktabs}
\usepackage[inline]{enumitem}
\usepackage{graphicx}
\usepackage{color}

\usepackage{listings}
\newsavebox\mypostbreak
\savebox\mypostbreak{\raisebox{0ex}[0ex][0ex]{\ensuremath{\color{red}\hookrightarrow\space}}}
\definecolor{delim}{RGB}{20,105,176}
\definecolor{numb}{RGB}{106, 109, 32}
\definecolor{string}{rgb}{0.64,0.08,0.08}
\lstdefinelanguage{json}{
    basicstyle=\normalfont\ttfamily,
    breaklines=true,
    literate=
         *{0}{{{\color{numb}0}}}{1}
          {1}{{{\color{numb}1}}}{1}
          {2}{{{\color{numb}2}}}{1}
          {3}{{{\color{numb}3}}}{1}
          {4}{{{\color{numb}4}}}{1}
          {5}{{{\color{numb}5}}}{1}
          {6}{{{\color{numb}6}}}{1}
          {7}{{{\color{numb}7}}}{1}
          {8}{{{\color{numb}8}}}{1}
          {9}{{{\color{numb}9}}}{1}
          {\{}{{{\color{delim}{\{}}}}{1}
          {\}}{{{\color{delim}{\}}}}}{1}
          {[}{{{\color{delim}{[}}}}{1}
          {]}{{{\color{delim}{]}}}}{1},
}

\newcommand{\BibTeX}{B\kern-.05em{\sc i\kern-.025em b}\kern-.08em\TeX}

\begin{document}


\begin{frontmatter}


\paperid{123} 


\title{ A Hate Speech Moderated Chat Application: Use Case for GDPR and DSA Compliance}


\author[A,B]{\fnms{Jan}~\snm{Fillies}\orcid{0000-0002-2997-4656}\thanks{Corresponding Author. Email: fillies@infai.org}}
\author[A,C]{\fnms{Theodoros}~\snm{Mitsikas}\orcid{0000-0002-7570-3603}}

\author[D]{\fnms{Ralph}~\snm{Schäfermeier}\orcid{0000-0002-4349-6726}}
\author[A,B,E]{\fnms{Adrian}~\snm{Paschke}\orcid{0000-0003-3156-9040}}

 \address[A]{Institut für Angewandte Informatik, Leipzig, Germany}
 \address[B]{Freie Universität Berlin, Berlin, Germany}
 \address[C]{National Technical University of Athens, Zografou, Greece}
\address[D]{Leipzig University, Leipzig, Germany}
 \address[E]{Fraunhofer-Institut für Offene Kommunikationssysteme, Berlin, Germany}


\begin{abstract}
The detection of hate speech or toxic content online is a complex and sensitive issue. While the identification itself is highly dependent on the context of the situation, sensitive personal attributes such as age, language, and nationality are rarely available due to privacy concerns. Additionally, platforms struggle with a wide range of local jurisdictions regarding online hate speech and the evaluation of content based on their internal ethical norms. This research presents a novel approach that demonstrates a GDPR-compliant application capable of implementing legal and ethical reasoning into the content moderation process. The application increases the explainability of moderation decisions by utilizing user information. Two use cases fundamental to online communication are presented and implemented using technologies such as GPT-3.5, Solid Pods, and the rule language Prova. The first use case demonstrates the scenario of a platform aiming to protect adolescents from potentially harmful content by limiting the ability to post certain content when minors are present. The second use case aims to identify and counter problematic statements online by providing counter hate speech. The counter hate speech is generated using personal attributes to appeal to the user. This research lays the groundwork for future DSA compliance of online platforms. The work proposes a novel approach to reason within different legal and ethical definitions of hate speech and plan the fitting counter hate speech. Overall, the platform provides a fitted protection to users and a more explainable and individualized response. The hate speech detection service, the chat platform, and the reasoning in Prova are discussed, and the potential benefits for content moderation and algorithmic hate speech detection are outlined. A selection of important aspects for DSA compliance is outlined.

\end{abstract}

\end{frontmatter}


\section{Introduction}
\label{sec:introduction}

``Content moderation is the organized practice of screening user-generated content'' \cite{roberts2017content}. It is a highly sensitive issue that directly influences a person's online safety. 

The Digital Services Act (DSA) was adopted in October 2022 and has been applicable since February 2024. The goal of the DSA is to define a comprehensive framework to counteract the dissemination of illegal and problematic content. It proposes a layered framework that defines different rules for different scopes. For a detailed view, refer to \citet{husovec2022digital}. One of the key problems of the DSA is that it does not harmonize what content or behavior is considered illegal; this remains under the sovereignty of the member states \cite{husovec2022digital}.

\citet{husovec2022digital} further states that a crucial aspect of the DSA is that online platforms accessible to minors must implement measures to ensure a high level of privacy, safety, and security. Additionally, they note that hosting providers must conduct fair content moderation. Uploaders are entitled to an explanation for the providers’ actions, whether the action is based on legal violations or terms of use violations. These aspects are not applicable to all platforms and do not exhaustively cover everything that needs to be fulfilled, but they are key aspects of the new regulation.

The DSA Act lays the groundwork for any moderation system and significantly influences the future of online communication.

Another important legislation is the European GDPR (General Data Protection Regulation) which was introduced in 2016 to set guidelines for personal data protection. These guidelines cover all major areas of life, setting standards and rules for handling personal data. A key aspect of GDPR is the ability to consent to and revoke consent for data processing. In the case of data processing through a data controller, it is necessary for the user (data subject) to be able to consent to or reject the processing (European Commission, 2016, Article 6). Furthermore, the user must also have the right to access all collected personal data in a machine-readable format and be able to transfer it to a different data controller (right to data portability) (European Commission, 2016, Article 20).

In the sensitive field of online content moderation, data privacy is highly important. On the one hand, having access to certain personal information of stakeholders in an online conversation can enable an unbiased and reliable system to be more precise and fair in performing automated moderation, as well as providing effective countermeasures against hate speech. On the other hand, this personal information can be very sensitive, necessitating strict guidelines on how to handle it and in what context. Managing and moderating online written content requires robust procedures that adhere to GDPR regulations, ensuring that user data is protected while maintaining a safe environment. Online platforms are balancing between internal community guidelines, and the jurisdictions covering the users and organizations, these are drivers of the complexity of the problem.

If a moderation system needs to use personal data or online communication in general, it must comply with these regulations. Therefore, it should be able to implement different levels of policies and handle the range of complexities occurring within the system. This starts with the simple execution of ground rules and extends to instances where rules are overridden, establishing a hierarchy that prioritizes some rules over others depending on the situation.

This research establishes a system for GDPR-compliant content monitoring capable of representing non-monotonic states and fulfilling the mentioned key aspects of the DSA. To this end, we present two different use cases (UC) for hate speech detection in online chatrooms modeled using the rule language Prova, Solid Pods, GPT-3.5 based hate speech detection and personalized counter hate speech generation. The use cases include typical stakeholders such as users, the platform, and data controllers. Access control and moderation are realized using concepts such as user consent, the purpose of access, and the role of the party requesting access. Prova and Solid have been used in various domains and similar contexts, demonstrating their versatility and effectiveness in applications requiring compliance with data protection regulations. The system is combined and demonstrated in a prototype implementing GDPR-compliant data sharing for content monitoring, using personal attributes during moderation and automatic counter hate speech generation. The prototype is available for testing. \footnote{http://81.169.159.230:7000/}

The research established the following main research objectives:

\begin{enumerate}
    \item A legal and ethical reasoning system for content moderation.
    \item Counter hate speech generation based on personal attributes.
    \item A chat platform for GDPR and DSA compliant  content moderation.
\end{enumerate}

The paper is organized as follows: Sections  \ref{sec:relatedwork} presents related work. Section \ref{sec:technicalpreliminaries} details the technical preliminaries such as Prova and Solid. Followed by Section \ref{sec:usecase} outlining both use cases.
Section \ref{sec:implementation} describes the implementation of the prototype. In Section \ref{sec:discussion} the work and ethical considerations are discussed. Followed lastly be the conclusion and future work in Section \ref{sec:conclusion}.


\section{Related Work}
\label{sec:relatedwork}
In the field of hate speech detection, historically, transformer-based architectures \citep{mozafari2020bert} and fine-tuning of transformer-based models \citep{filliesbigdata}, specifically BERT \citep{devlin2019bert}, have yielded better performance compared to traditional machine learning models \citep{liu2019nuli,caselli2021hatebert,mathew2021hatexplain}. In recent years, pre-trained large language models have gained traction \citep{kim-etal-2023-conprompt, kumarage2024harnessing} due to their performance and simple setup. LLMs have also proven efficient in the field of counter hate speech generation \citep{wang2023evaluating}, with the capability to effectively generate personalized counter hate speech \citep{doganc-markov-2023-generic}.

In the area of compliance checking, many different works have been established in recent years \cite{satoh2010proleg, governatori2011designing,recomp_challenge}. \citet{satoh2010proleg} proposed a legal reasoning system for decision-making by judges under incomplete information. \citet{recomp_challenge} established a compliance mechanism for AI agent planning in a multi-agent setting. \citet{goossens2023gpt} showed the possibilities of using GPT-3 in decision logic modeling, and \citet{hayashi2023online} presented a planning method for legal and ethical norms.
 
Two main works in the field could be established. Firstly, \citet{schafermeier2022modeling} proposed a distributed data wallet use case that is GDPR-compliant, comparing two different approaches by applying AspectOWL and Prova for the modeling and implementation. AspectOWL is a monotonic contextualized ontology language that focuses on the representation of dynamic state transitions and knowledge retention by wrapping parts of the ontology in isolated contexts. In contrast, Prova handles state transitions at runtime using non-monotonic state transition semantics. They analyzed two use cases: one providing a personalized search and the other outlining the process of sharing pictures via a wallet-enabled sharing app. Both use cases were implemented and evaluated on aspects such as human and machine-readability, manageability, and the use of open standard technology. One of the findings was that AspectOWL is suitable for specifying the ontological domain model, while Prova is a more practical approach for real-world applications, including the interaction between involved parties.

The second research by \citet{mitsikasmodeling} presents a medical data access use case compliant with GDPR legal rules, also implemented using Prova. It demonstrates a scenario of a patient consenting to medical data sharing. The data is used for a specific purpose, and cases were considered where the typical rules are overridden, thereby adjusting the access rights.

This current research uses modern algorithms for hate speech detection and builds upon the works by \citet{schafermeier2022modeling} and \citet{mitsikasmodeling}, but also others, as \citet{recomp_challenge}.  It follows existing research into designing a GDPR-compliant application, also choosing Prova for development due to its practicality and scalability. This research advances the field with two new highly important use cases and incorporates key aspects of the DSA legislation.

\section{Technical Preliminaries}
\label{sec:technicalpreliminaries}

\subsection{Prova}
\label{subsec:prova}
Prova is both a (Semantic) Web rule language and a distributed (Semantic) Web rule engine. It supports reaction rule based workflows, event processing, and reactive agent programming. It integrates Java scripting with derivation and reaction rules, and message exchange with various communication frameworks~\cite{prova,kober2022modeling,Paschke2011}. 

Syntactically, Prova builds upon the ISO Prolog syntax and extends it, notably with the integration of Java objects, typed variables, F-Logic-style slots, and SPARQL and SQL queries. 
Slotted terms in Prova are implemented using the arrow expression syntax `\texttt{->}'as in RIF and RuleML, and can be used as sole arguments of predicates. They correspond to a Java HashMap, with the keys limited to Stings~\cite{kozlenkov2010prova}.

Semantically, Prova provides the expressiveness of serial Horn logic with a linear resolution for extended logic programs (SLE resolution)~\cite{PASCHKE2008187}, extending the linear SLDNF resolution with goal memoization and loop prevention. Negation as failure support in the rule body can be added to a rulebase by implementing it using the cut-fail test as follows:
{\small
\begin{lstlisting}
not(A) :- derive(A), !, fail().
not(_).
\end{lstlisting}
}

Prova implements an inference extension called literal \emph{guards}, specified using brackets. By using guards, we can ensure that during unification, even if the target rule matches the source literal, further evaluation is delayed unless a guard condition evaluates to true. Guards can include arbitrary lists of Prova literals including Java calls, arithmetic expressions, relations, and even the cut operator. Prova guards play even a more important role in message and event processing, as they allow the received messages to be examined before they are irrevocably accepted. The guards are tested right after pattern matching but before a message is fully accepted, so that the net effect of the guard is to serve as an extension of pattern matching for literals~\cite{kozlenkov2010prova,paschkereaction}. 
\subsection{Solid}
\label{subsec:aspectowl}

The Solid platform, first introduced by \citet{sambra2016solid}, is a decentralized platform using W3C standards to create social applications based on linked data approaches. Linked data is a form of data interlinked with each other and accessible via semantic queries \cite{ramachandran2020towards}. As described by \citet{mansour2016demonstration}, following the concept of Solid, the data of each user is stored independently of the sources that created it, the data broker, and the end data consumer. Each user owns and manages their own personal online datastore (Pod) where all personal data is stored. A user is not limited to one Pod or one hosting provider, as they can self-host their Pods or choose between different hosting providers.

Applications that want to work with and access the data use protocols based on W3C standards. \citet{mansour2016demonstration} further states that a decentralized authentication and access control mechanism lays the groundwork for strong privacy protection. The decentralized architecture allows applications to access the user data independent of the hosting option, while users have full control over and access to their data at any point, with the possibility to switch providers or withdraw consent for sharing data.


\begin{figure}[!t]
\centering
\includegraphics[width=3.5in]{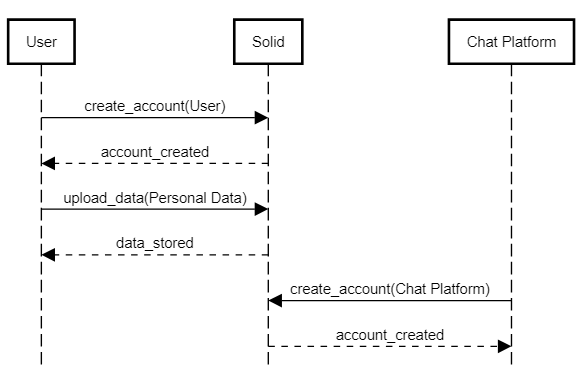}
\caption{Sequence diagram of the initial account creation steps every user has to do in both use cases. }
\label{sequence_create_account}
\end{figure}

\begin{figure}[!t]
\centering
\includegraphics[width=2.6in]{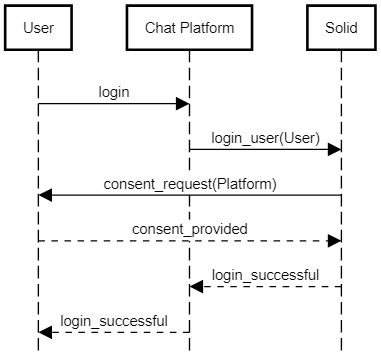}
\caption{Sequence diagram of the steps done to join a chatroom. Every user has to do these steps in both use cases. }
\label{sequence_join_chatroom}
\end{figure}

\begin{figure}[!t]
\centering
\includegraphics[width=3.3in]{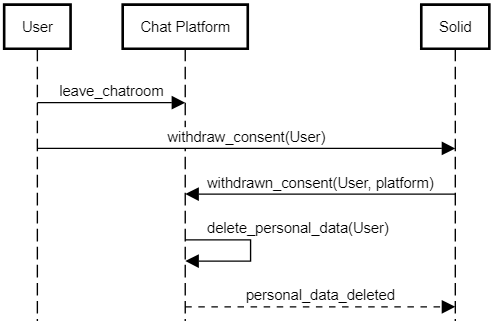}
\caption{Sequence diagram of leaving a chatroom. Steps every user has to do in both use cases. }
\label{sequence_leave_chatroom}
\end{figure}

\section{Use Cases}
\label{sec:usecase}

Two data wallet use cases are described in terms of interaction sequences and data exchange between the different parties involved. The use cases involve data wallet owners sharing personal data using relaying parties that provide specialized applications, such as an Age-based Content Moderation application and a Hate Speech Classification combined with a Contextualized Semantic Counter Narrative Generation. 

\subsection{General Steps}
\label{subsec:general}

Certain steps apply to both uses cases. Figure \ref{sequence_create_account} refers to the initial steps a user and the platform has to perform to create and register with the Solid platform. The solid platform represents the data controller in this setting, storing the personal data and managing its access. Figure \ref{sequence_join_chatroom} depicts the process of a user login into the chatroom, providing consent for accessing personal data to the solid instance, and finally joining a specific chatroom. Lastly, Figure \ref{sequence_leave_chatroom} is the process of leaving the chatroom, withdrawing the consent to user the personal data. In all figures, the requests are represented by solid arrows and the responses by dotted arrows.

\begin{figure}[!t]
\centering
\includegraphics[width=3.3in]{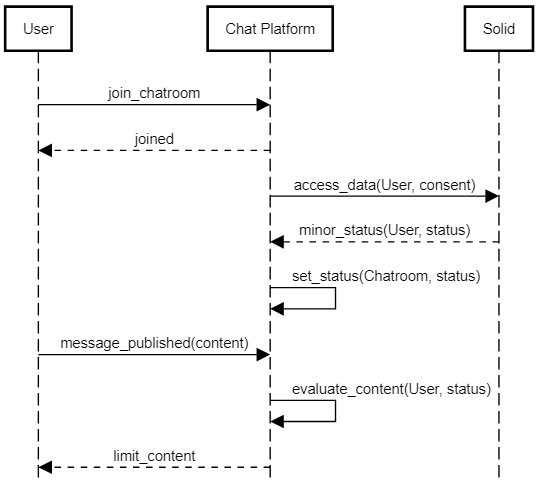}
\caption{Sequence diagram of the UC 1. }
\label{fig:uc1}
\end{figure}

\subsection{Use Case 1: Age-based Content Moderation}
\label{subsec:usecase1}

The following use case outlines the scenario in which a platform needs to adjust the visibility of certain content (e.g., highly offensive content) as soon as adolescents enter their communication platform. The user and platform both need to be logged in to the data controller, and consent must be granted.

The Use Case: A minor (age 14) joins a chatroom. In Germany, at the age of 14 and younger, an individual is considered a child. The platform made the internal decision to protect the child by limiting all highly toxic statements (e.g., Holocaust denial) posted to the chat during the presence of a minor.

\subsubsection{Interaction sequences and data exchange:}

See Figure \ref{fig:uc1} for the sequence diagram without the login, logout, or account creation. 

\begin{itemize}
    \item[--] The user logs into the chat platform and provides consent for their data to be accessed and joins a chatroom.
    \item[--] The chat platform requests the data controller to provide information if the user is a minor based on the country of origin and the personal age of the user.
    \item[--] Whenever a message is sent, as long as the user is present in the chatroom, the platform can limit the posted content.
    \item[--] When the user leaves the chatroom, withdrawing their consent for the data to be accessed, the chat is opened up for content suitable for adults.
\end{itemize}

\subsection{Use Case 2: Contextualized Semantic Hate Speech Classification Combined with Counter Narrative Generation}
\label{subsec:usecase2}

This use case focuses on the moment an adult user of an online platform writes a problematic statement, such as denying the Holocaust, to other adults. In this setting, the platform needs to make multiple decisions. Firstly, is the message legal for the person to publish here? Secondly, is the statement against its internal guidelines? Thirdly, how to best address the statement.

After the message is classified as denying the Holocaust, to make a fitting legal decision, personal information such as the location of the user is needed. This information is obtained from the data controller. Now the statement can be evaluated against legal and ethical guidelines using a compliance check. Lastly, based on the personal information, a contextualized counter hate speech can be generated.

The Use Case: On a chat platform, a US citizen from California posts a statement denying the Holocaust. The platform can evaluate it based on the personal data of the user. In the US, this statement is covered under freedom of speech, making it legal for him to post. However, due to their internal guidelines, the platform still decides against the publication of the content. This reasoning is integrated into the created counter hate speech in English, explaining the reason for the message to be classified as problematic within the cultural context of America.

In the same session, a Greek user from Delphi publishes a statement also denying the Holocaust. The platform again evaluates it based on the personal data of the user. In Greece, it is not legal to deny the Holocaust. The platform therefore blocks the content from being posted and integrates this into the created counter hate speech in Greek, explaining the reason for the message to be classified as problematic within the cultural context of Greece.

\subsubsection{Interaction sequences and data exchange:}

\begin{figure}[!t]
\centering
\includegraphics[width=3.3in]{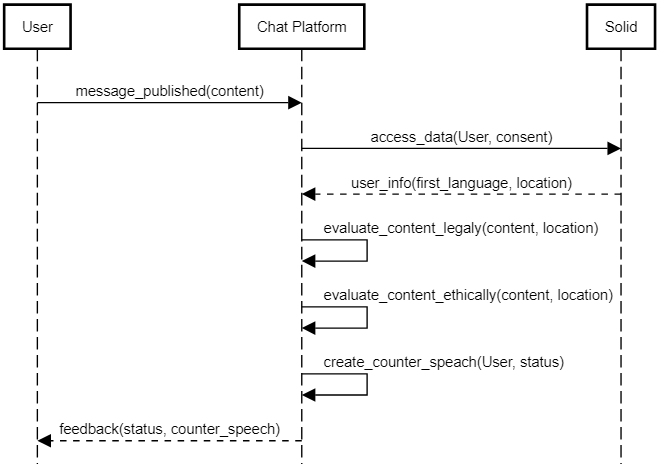}
\caption{Sequence diagram of the UC 2. }
\label{figure:uc2}
\end{figure}

See Figure \ref{figure:uc2} for the sequence diagram without the login, logout, or account creation. 

\begin{itemize}
    \item[--] A harmful user posts a statement denying the existence of the Holocaust.
    \item[--] The platform requests information from the data provider to determine if the statement violates their guidelines or local laws regulate such statements.
    \item[--] Based on the input, the platform provides feedback that the statement is problematic not only based on internal guidelines but also due to local jurisdiction.
    \item[--] The platform requests the first language and cultural background of the user from the data controller.
    \item[--] With this information, the platform delivers understandable counter hate speech and provides context as to why the post was problematic. 

\end{itemize}

\section{Implementation}
\label{sec:implementation}

\subsection{Architecture}
\label{subsec:architecture}
\begin{figure}[!t]
\centering
\includegraphics[width=3.3in]{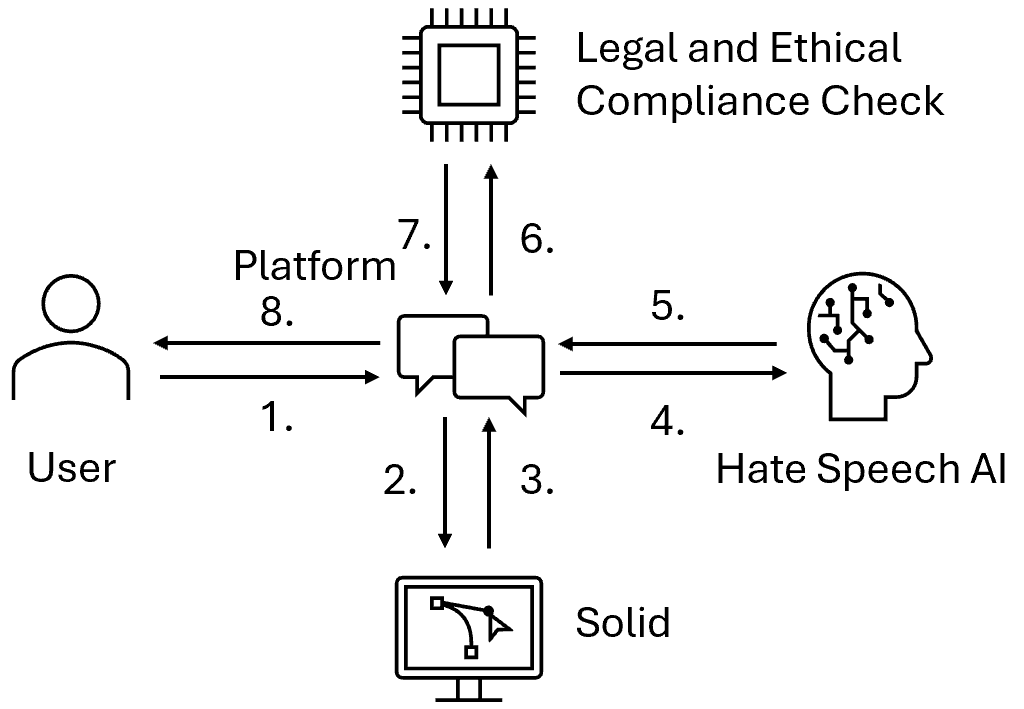}
\caption{First rough architecture overview. }
\label{architecturefig}
\end{figure}

As depicted in Figure \ref{architecturefig}, the prototype integrates the chat platform, the data controller (Solid), a hate speech detection service, and a Compliance Check implemented with Prova. All services are necessary to ensure safe and GDPR-compliant communication.

The chat platform serves as the interface for the user, supporting real-time messaging, and is designed to handle concurrent users. If the user sends a message to the platform (1. in Figure \ref{architecturefig}), the platform can request personal data from the user via the data controller, implemented as a Solid application (2.). After the controller provides the information (3.), the platform can forward the personal information and the original message to the hate speech API (4.), which evaluates the content for hate speech, such as Holocaust denial, and generates personalized counter hate speech (5.). Based on the classification result, the platform can then request the Compliance Check (6.) to determine if the message violates legal or ethical standards (7.). Based on all responses, the platform can act accordingly and interact with the original message (8.).  

\subsection{Solid}
\label{subsec:impl_solid}

The chat application has been implemented as a React\footnote{\url{https://react.dev}} application. For the authentication and communication with the Solid platform, Inrupt's JavaScript Client Libraries and the React SDK\footnote{\url{https://docs.inrupt.com/developer-tools/}} were used.

User data relevant to the use of the chat are stored in the user's profile (which is a standard Solid dataset and can be assumed to exist for every Solid user).
The user's name, age and location of origin are stored as RDF triples using the vCard\footnote{\url{https://www.w3.org/TR/vcard-rdf/}} and FOAF\footnote{\url{http://xmlns.com/foaf/spec/}} vocabularies, respectively. User data retrieved from the Solid Pod is retained in the chat application for the duration of a chat session only. No personal data is permanently stored outside of the user's Solid Pod.

\subsection{Chat}
\label{subsec:chat}

The chat application has been implemented using components from the Chat UI Kit\footnote{\url{https://chatscope.io}}.
The public demonstrator instance comes preconfigured with four chatrooms, each of them containing one virtual chat partner with different age and location of origin.

As soon as a new chat message is sent to any of the chatrooms, the text of the message is forwarded to the hate speech detection endpoint where it is being analyzed for hateful content (see also Section \ref{subsec:hatespeechdetection}).
The hate speech detection endpoint returns information about which kind of hate speech was detected (if any) and, if applicable, a numerical score ranging over 1-5 indicating the severity of the hate speech.

If the outcome is positive, a request containing the hate speech analysis result, age and location of the hateful comment's originator, as well as information about whether minors are present in the chatroom are sent to the legal and ethical compliance checker.
The latter one decides whether the given instance of hate speech constitutes a violation against ethical norms (such as the chat service's community guidelines) or legal norms (considering the user's location) or both.
See Section \ref{subsec:impl_prova} for details on the compliance checking process.

In case of the presence of hate speech with ethical but without legal relevance, the hate speech mitigation endpoint is requested to generate appropriate counter speech.

\begin{figure}[!t]
\centering
\includegraphics[width=\columnwidth]{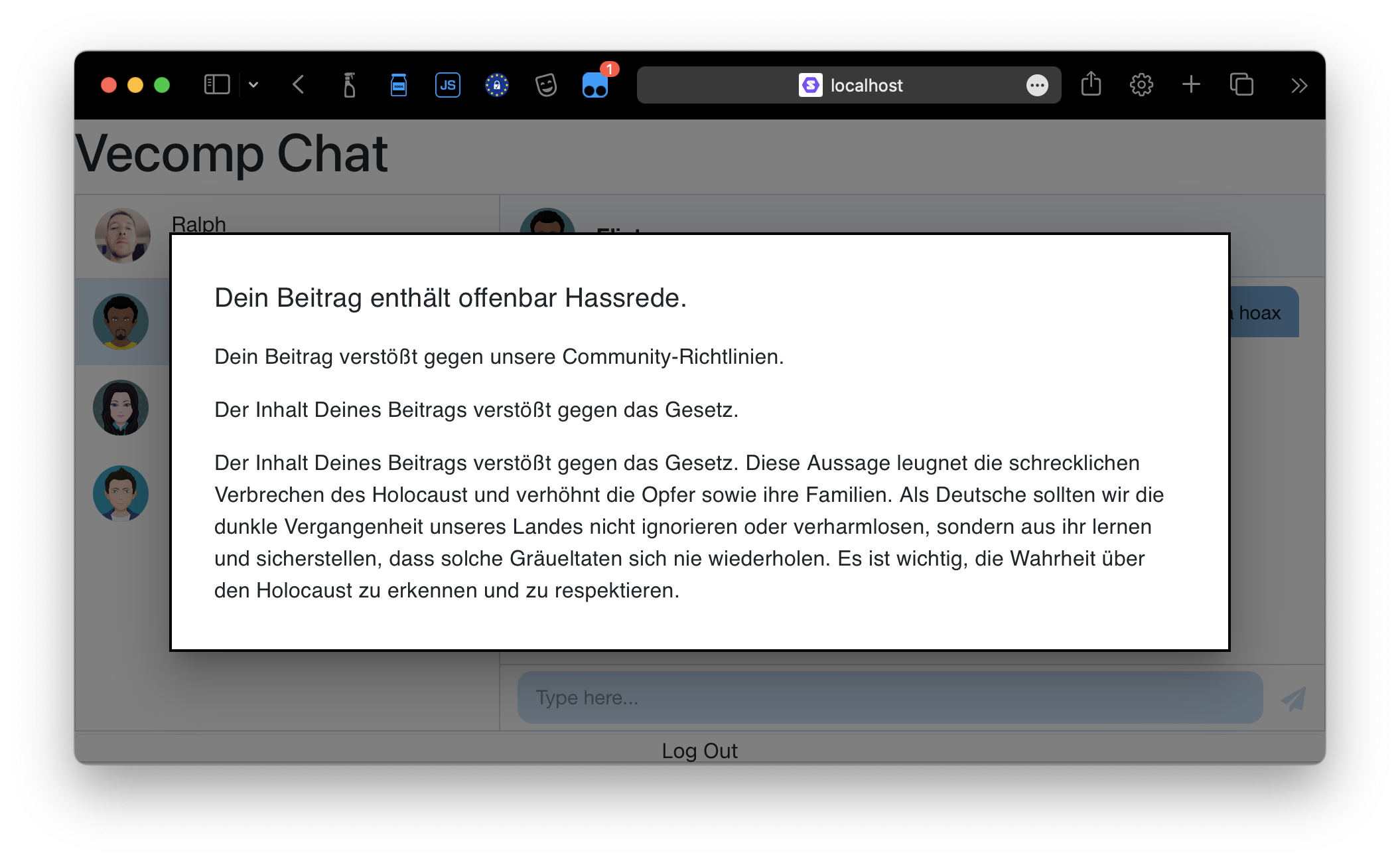}
\vspace{-.8cm}
\caption{Screenshot of the chat application displaying a warning message in German that a user's post contains hate speech violating community guidelines and national law as well as a personalized counter hate speech message explaining the decision. }
\label{chatscreenshotfig}
\end{figure}

Depending on the outcome, the chat application suppresses the message and presents the harmfully acting user with a warning message containing details about the reason for the intervention and the counter speech message.
Figure \ref{chatscreenshotfig} shows a screenshot of the application displaying a warning message about the violation of ethical and legal rules by one of the users' posts.

\subsection{Hate Speech Detection}
\label{subsec:hatespeechdetection}
As shown by \citet{kumarage2024harnessing}, OpenAI’s LLMs are able to achieve stellar results in the tasks of hate speech detection. To build a simple but effective classifier, this research utilizes the OpenAI API with the GPT-3.5-turbo model as the foundation for hate speech detection. Due to its relatively recent release, size, economic considerations, and performance on academic benchmarks \cite{kumarage2024harnessing}.

\citet{Ekin2023} suggests different approaches to prompt engineering. In his work, basic and advanced strategies are outlined. Due to hate speech detection not being the main focus of this work, only basic methods are used, applying templates and iterative testing and refinement. The advanced strategies would involve utilizing temperature and token control, prompt chaining, and adapting prompts. For this approach, templates were designed and then iteratively tested and refined until the results were satisfactory to the group of researchers.

Example prompt: ``Analyze the following text for hate speech and respond with `hate' or `no-hate'. Provide a level of hate from 1-5, and indicate if Holocaust denial is present.''

In the field of counter hate speech generation, \citet{wang2023evaluating} shows the strong capabilities of automatic counter hate speech generation utilizing GPT-3.5. They highlighted these capabilities while also raising concerns about the high need for well-designed prompts. \citet{doganc-markov-2023-generic} demonstrate the possibility of creating high-quality counter hate speech statements by including personalized aspects. Based on these findings, this research again utilized templates and iterative testing and refinement to design the counter-narrative generation.

Example prompt: ``Generate a counter speech for the following text. You are an observer. The response should be between 50-100 words, contextualizing the problematic statements for someone of \{national\_origin\} origin and providing the counter speech in \{language\}.''

The hate speech detection and counter hate speech generation service is provided via a Swagger-based API interface. The underlying code is Python. The cost per call to the OpenAI API is \$0.50 per 1M tokens\footnote{https://openai.com/api/pricing/}.

\subsection{Compliance Checking with Prova}
\label{subsec:impl_prova}
The compliance checking mechanism, implemented in Prova, evaluates possible violations of legal and ethical standards. It functions as a service implemented in Prova and Java, that accepts HTTP requests of the following set of parameters: {\tt user\_location}, {\tt user\_age}, {\tt chat\_context}, {\tt hate\_speech\_score}, and {\tt hol}. The parameter {\tt hol} characterizes the user message about the presence (or absence) of Holocaust denial, while the parameter {\tt chat\_context} denotes the general context of the chat, for example, if it is a chat where minors are participating. While not all parameters are used for this use case, they are included, aiming at future expansion.

The parameters (and their values) are converted to Prova slots (pairs of key and value), and are passed as messages to the two rulebases that perform, in turn, the legal and the ethical check. 

First, the legal checker is invoked, using a subset of the slots, to check for potential legal violations of the user message. This depends on the message content, as well as the user location. The countries where Holocaust denial is a legal violation is provided through the {\tt illCountry} predicate. In this case, three rule variants exist: 
\begin{enumerate*}
    \item if the user message denies the Holocaust and the user location is in a country where Holocaust denial is illegal, 
    \item if the user message  denies the Holocaust and the user location is not in a country where Holocaust denial is illegal, 
    \item if the user message does not deny the Holocaust
\end{enumerate*}

\begin{lstlisting}[language=Prolog]
legalChecker() :-
     rcvMult(X,P,F,executionRequest, {hol->hol_denial,user_location->L}) [illCountry(L)],
     spawn(X,$Service,result, ["legal_violation",      "Holocaust Denial"]),
     spawn(X,$Service,resume,[]).

legalChecker() :-
     rcvMult(X,P,F,executionRequest, {hol->hol_denial,user_location->L}) [not(illCountry(L))],
     spawn(X,$Service,resume,[]).

legalChecker() :-
     rcvMult(X,P,F,executionRequest,{hol->H}) [not_equal(H,hol_denial)],
     spawn(X,$Service,resume,[]).
\end{lstlisting}
As shown above, the legal checker rulebase first selects the relevant messages through pattern matching over the slots (e.\,g., {\tt user\_location->L)}, and then irrevocably accepts them if the guard (e.\,g., \verb+[not(illCountry(L))]+) is satisfied, proceeding with calling outside Java methods that update the service's answer. In particular,  \verb+spawn(X,$Service,result,["...", "..."])+ calls the Java method {\tt result(String, String)}, which updates the answer with a kind of violation ({\tt legal\_violation}, or {\tt ethical\_violation}), while \verb+spawn(X,$Service,resume,[])+ invokes the method {\tt resume()} that invokes a {\tt notifyAll()} Java call. The latter is implemented for performance reasons, resuming the execution of the main Java thread (as Prova runs on different threads) as soon as Prova updates the answer. 
The third rule exists for the performance reasons mentioned above.

After the legal check and the potential update of the response, the ethical checker is called, to finalize the response. It contains analogous rules for the ethical checking, where the location of the user is not checked (Holocaust denial is unethical regardless of the user location), as well as a check for other ethical violations denoted by the parameter {\tt hate\_speech\_score}. 

The final response is provided in JSON form, for example
\begin{lstlisting}[language=json]
{
   "response":{
      "legal_violation":{
         "reason":"Holocaust Denial"
      },
      "ethical_violation":{
         "reason":"Holocaust Denial",
         "score":5
      }
   }
}
\end{lstlisting}

\section{Discussion and Ethical Consideration}
\label{sec:discussion}

The created platform establishes the primary requirements of the GDPR (see Section \ref{sec:introduction}) by separating the individual components into data subjects, data controllers, data processors, and an independent identification service. It provides users with a clear option to not only consent to their data sharing but also revoke access at any time. Additionally, by storing the shared data in their individual Solid Pods, which are provider-independent, users have full control over the type of data shared and all current access rights. While there are more aspects to the GDPR, this research covers the main parts and adopts a similar approach to other GDPR-compliant applications, such as in \citep{schafermeier2022modeling}.

Regarding the DSA, three key aspects were introduced in Section~\ref{sec:introduction}. Firstly, the DSA does not harmonize what constitutes illegal content. The introduced compliance checker can include different legal and ethical definitions of hate speech. It is important that personal information needed to make these decisions can be shared securely and legally within this system, addressing a major open problem of the DSA. While the focus of this application was not on modeling all legislation regarding illegal content or ethical understanding of hate speech, the use cases were designed with a clear, small scope to show that the architecture and application can handle such a complex setting and can now be extrapolated and generalized to a broader spectrum.

Secondly, based on the DSA, online platforms that involve minors must take measures to ensure a high level of privacy, safety, and security. The proposed platform demonstrates this in the first use case, showing that the system can account for the presence of minors and adapt its behavior accordingly. A high level of data security is universally fulfilled with the proposed Solid infrastructure and GDPR-compliant structuring. Similar to the second point, the application does not introduce a complete and absolute solution on how to handle minors within a social platform but rather shows a way to generally provide privacy, safety, and security. This concept must now be adapted and fitted to more advanced features.

Thirdly, hosting providers must conduct fair content moderation. Users must be informed about moderation decisions, for example, whether the action is based on legal violations or violations of the terms of use. This key aspect is covered as shown in use case two. Here, the user is not only informed if their content was removed based on legal or ethical concerns but also receives a personalized response in their native language and with consideration of their social context, provided in natural language. Furthermore, it is important to mention that personal information is used in the ethical compliance checker to identify if local law was broken, making it a context-based hate speech detection system.

Content moderation is always a fine line between protecting people from online harm and limiting the ability to express oneself freely. This research developed a tool that supports content moderation, emphasizing that the researchers advocate for human-in-the-loop moderation approaches. It is possible that the LLM will make classification mistakes, this can only be ultimately solved by a human in the loop. Since both are prone to error, a mixed approach seems the best solution. This research is not intended to be viewed as a fully automatic solution. Furthermore, the proposed solution can include contextual personal information to make more informed legal and ethical decisions, distinguishing it from existing solutions that can handle either a mixture or just one type of information.

Using Large Language Models (LLMs) to generate counter speech based on personal attributes is a very young research discipline. While initial studies show that it is possible, no large-scale testing on the reliability or ethical aspects has been conducted. In this work, only language and country of origin are used for the generation process, both with explicit, always revocable consent. The ``language'' attribute is necessary to address the person in a format they understand best. The attribute ``country of origin'' could be more problematic, as it may result in unfitting counter hate speech. However, this risk is minimal considering that the LLMs used have safeguards in place to prevent discrimination and hate in their responses.

By introducing a legal and ethical compliance check, this research ensures a clear distinction between legal and ethical considerations, while also protecting legal and ethical statements. This work is in the public interest, focusing primarily on legislation such as the GDPR and DSA. The risk of sharing sensitive personal data is more manageable due to the full knowledge, control, and consent of the person sharing the data, in contrast to other standard data-sharing practices. 


\section{Conclusion and Future Work}
\label{sec:conclusion}

The research outlines a GDPR-compliant application in the field of hate speech moderation. It lays the groundwork for key aspects of future DSA compliance. Two new use cases are introduced and implemented using Python, Prova, and Java. The first use case covers a key requirement regarding the protection of minors online introduced by the DSA. The second one shows the possibility of fair content moderation.

The architecture consists of four components: a platform that serves as the interface to the user and manages communication with the other tools, a Solid instance for access, permission, and storage of personalized user data, and identification of the users, an API that detects hate speech and Holocaust denial in text and generates counter hate speech based on personal attributes, and the Legal and Ethical Compliance Checker that evaluates specific instances based on Prova implementation for different legal and ethical scenarios. The compliance checker is able to contain formalized legislative rules and, based on the country of origin, identify if the given statement is considered illegal in a certain country (demonstrated in the case of Holocaust denial in Europe).

The architecture is clearly split into the different stakeholders required by the GDPR, and the required rights to consent and withdraw consent to data sharing are fulfilled.

In general, the application is the first known prototype to address these challenges arising with the DSA and GDPR in the context of content moderation. It provides a working demonstrator that shows the applicability and functionality of the proposed architecture and solutions.

In the future, more legal definitions need to be included in the compliance checker, extending on the one trial implementation. Furthermore, one of the strong suits that need to be explored is the possibility of using personal information directly in the LLM to identify context-based hate speech. The architecture could be expanded to also include a human-in-the-loop aspect for better safety and quality control. The proposed system needs to be further evaluated regarding its usability but also scalability and performance aspects. Lastly, the application could be expanded upon in the sense of other DSA aspects.







\begin{ack}
This work has been partially funded by the  Deutsche Forschungsgemeinschaft (DFG, German Research Foundation) project RECOMP (DFG – GZ: PA 1820/5-1) and by the German Federal Ministry of Education and Research, project ``Terminology and Ontology-Based Phenotyping (TOP)'' (grant number: 01ZZ2018).
\end{ack}



\bibliography{main}

\begin{thebibliography}{29}
\providecommand{\natexlab}[1]{#1}
\providecommand{\url}[1]{\texttt{#1}}
\expandafter\ifx\csname urlstyle\endcsname\relax
  \providecommand{\doi}[1]{doi: #1}\else
  \providecommand{\doi}{doi: \begingroup \urlstyle{rm}\Url}\fi

\bibitem[Caselli et~al.(2021)Caselli, Basile, Mitrovi{\'c}, and Granitzer]{caselli2021hatebert}
T.~Caselli, V.~Basile, J.~Mitrovi{\'c}, and M.~Granitzer.
\newblock {H}ate{BERT}: Retraining {BERT} for abusive language detection in {E}nglish.
\newblock In A.~Mostafazadeh~Davani, D.~Kiela, M.~Lambert, B.~Vidgen, V.~Prabhakaran, and Z.~Waseem, editors, \emph{Proceedings of the 5th Workshop on Online Abuse and Harms (WOAH 2021)}, pages 17--25, Online, Aug. 2021. Association for Computational Linguistics.
\newblock \doi{10.18653/v1/2021.woah-1.3}.
\newblock URL \url{https://aclanthology.org/2021.woah-1.3}.

\bibitem[Devlin et~al.(2019)Devlin, Chang, Lee, and Toutanova]{devlin2019bert}
J.~Devlin, M.-W. Chang, K.~Lee, and K.~Toutanova.
\newblock {BERT}: Pre-training of deep bidirectional transformers for language understanding.
\newblock In J.~Burstein, C.~Doran, and T.~Solorio, editors, \emph{Proceedings of the 2019 Conference of the North {A}merican Chapter of the Association for Computational Linguistics: Human Language Technologies, Volume 1 (Long and Short Papers)}, pages 4171--4186, Minneapolis, Minnesota, June 2019. Association for Computational Linguistics.
\newblock \doi{10.18653/v1/N19-1423}.
\newblock URL \url{https://aclanthology.org/N19-1423}.

\bibitem[Do{\u{g}}an{\c{c}} and Markov(2023)]{doganc-markov-2023-generic}
M.~Do{\u{g}}an{\c{c}} and I.~Markov.
\newblock From generic to personalized: Investigating strategies for generating targeted counter narratives against hate speech.
\newblock In Y.-L. Chung, H.~Bonaldi, G.~Abercrombie, and M.~Guerini, editors, \emph{Proceedings of the 1st Workshop on CounterSpeech for Online Abuse (CS4OA)}, pages 1--12, Prague, Czechia, Sept. 2023. Association for Computational Linguistics.
\newblock URL \url{https://aclanthology.org/2023.cs4oa-1.1}.

\bibitem[Ekin(2023)]{Ekin2023}
S.~Ekin.
\newblock Prompt engineering for chatgpt: A quick guide to techniques, tips, and best practices, 05 2023.

\bibitem[Fillies et~al.(2023)Fillies, Hoffmann, and Paschke]{filliesbigdata}
J.~Fillies, M.~Hoffmann, and A.~Paschke.
\newblock Multilingual hate speech detection: Comparison of transfer learning methods to classify german, italian, and spanish posts.
\newblock In \emph{2023 IEEE International Conference on Big Data (BigData)}, pages 5503--5511, Los Alamitos, CA, USA, dec 2023. IEEE Computer Society.
\newblock \doi{10.1109/BigData59044.2023.10386244}.
\newblock URL \url{https://doi.ieeecomputersociety.org/10.1109/BigData59044.2023.10386244}.

\bibitem[Goossens et~al.(2023)Goossens, Vandevelde, Vanthienen, and Vennekens]{goossens2023gpt}
A.~Goossens, S.~Vandevelde, J.~Vanthienen, and J.~Vennekens.
\newblock {GPT-3} for decision logic modeling.
\newblock \emph{Proceedings of the 17th International Rule Challenge and 7th Doctoral Consortium@ RuleML+ RR 2023 co-located with 19th Reasoning Web Summer School (RW 2023) and 15th DecisionCAMP 2023 as part of Declarative AI 2023}, 3485:\penalty0 1--14, 2023.

\bibitem[Governatori et~al.(2011)Governatori, Olivieri, Scannapieco, and Cristani]{governatori2011designing}
G.~Governatori, F.~Olivieri, S.~Scannapieco, and M.~Cristani.
\newblock Designing for compliance: Norms and goals.
\newblock In \emph{International Workshop on Rules and Rule Markup Languages for the Semantic Web}, pages 282--297. Springer, 2011.

\bibitem[Hayashi and Satoh(2023)]{hayashi2023online}
H.~Hayashi and K.~Satoh.
\newblock Online htn planning for data transfer and utilization considering legal and ethical norms: Case study.
\newblock In \emph{ICAART (1)}, pages 154--164, 2023.

\bibitem[Hayashi et~al.(2023)Hayashi, Mitsikas, Taheri, Tsushima, Schäfermeier, Bourgne, Ganascia, Paschke, and Satoh]{recomp_challenge}
H.~Hayashi, T.~Mitsikas, Y.~Taheri, K.~Tsushima, R.~Schäfermeier, G.~Bourgne, J.-G. Ganascia, A.~Paschke, and K.~Satoh.
\newblock Multi-agent online planning architecture for real-time compliance.
\newblock In \emph{Proceedings of the 17th International Rule Challenge and 7th Doctoral Consortium \@ RuleML+RR 2023}, volume 3485. CEUR, 2023.
\newblock URL \url{https://ceur-ws.org/Vol-3485/}.

\bibitem[Husovec and Roche~Laguna(2022)]{husovec2022digital}
M.~Husovec and I.~Roche~Laguna.
\newblock Digital services act: A short primer.
\newblock \emph{Martin Husovec and Irene Roche Laguna, Principles of the Digital Services Act (Oxford University Press, Forthcoming 2023)}, 2022.

\bibitem[Kim et~al.(2023)Kim, Park, Namgoong, and Han]{kim-etal-2023-conprompt}
Y.~Kim, S.~Park, Y.~Namgoong, and Y.-S. Han.
\newblock {C}on{P}rompt: Pre-training a language model with machine-generated data for implicit hate speech detection.
\newblock In H.~Bouamor, J.~Pino, and K.~Bali, editors, \emph{Findings of the Association for Computational Linguistics: EMNLP 2023}, pages 10964--10980, Singapore, Dec. 2023. Association for Computational Linguistics.
\newblock URL \url{https://aclanthology.org/2023.findings-emnlp.731}.

\bibitem[Kober et~al.(2022)Kober, Robaldo, and Paschke]{kober2022modeling}
G.~Kober, L.~Robaldo, and A.~Paschke.
\newblock {Modeling Medical Guidelines by Prova and SHACL Accessing FHIR/RDF. Use Case: The Medical ABCDE Approach}.
\newblock In \emph{dHealth 2022}, pages 59--66. IOS Press, 2022.

\bibitem[Kozlenkov(2010)]{kozlenkov2010prova}
A.~Kozlenkov.
\newblock \emph{{Prova Rule Language version 3.0 User’s Guide}}, 2010.
\newblock URL \url{https://github.com/prova/prova/tree/master/doc}.

\bibitem[Kozlenkov et~al.(2006)Kozlenkov, Penaloza, Nigam, Royer, Dawelbait, and Schroeder]{prova}
A.~Kozlenkov, R.~Penaloza, V.~Nigam, L.~Royer, G.~Dawelbait, and M.~Schroeder.
\newblock {Prova: Rule-Based Java Scripting for Distributed Web Applications: A Case Study in Bioinformatics}.
\newblock In T.~Grust, H.~H{\"o}pfner, A.~Illarramendi, S.~Jablonski, M.~Mesiti, S.~M{\"u}ller, P.-L. Patranjan, K.-U. Sattler, M.~Spiliopoulou, and J.~Wijsen, editors, \emph{Current Trends in Database Technology -- EDBT 2006}, pages 899--908, Berlin, Heidelberg, 2006. Springer.
\newblock ISBN 978-3-540-46790-8.

\bibitem[Kumarage et~al.(2024)Kumarage, Bhattacharjee, and Garland]{kumarage2024harnessing}
T.~Kumarage, A.~Bhattacharjee, and J.~Garland.
\newblock Harnessing artificial intelligence to combat online hate: Exploring the challenges and opportunities of large language models in hate speech detection, 2024.

\bibitem[Liu et~al.(2019)Liu, Li, and Zou]{liu2019nuli}
P.~Liu, W.~Li, and L.~Zou.
\newblock {NULI} at {S}em{E}val-2019 task 6: Transfer learning for offensive language detection using bidirectional transformers.
\newblock In J.~May, E.~Shutova, A.~Herbelot, X.~Zhu, M.~Apidianaki, and S.~M. Mohammad, editors, \emph{Proceedings of the 13th International Workshop on Semantic Evaluation}, pages 87--91, Minneapolis, Minnesota, USA, June 2019. Association for Computational Linguistics.
\newblock \doi{10.18653/v1/S19-2011}.
\newblock URL \url{https://aclanthology.org/S19-2011}.

\bibitem[Mansour et~al.(2016)Mansour, Sambra, Hawke, Zereba, Capadisli, Ghanem, Aboulnaga, and Berners-Lee]{mansour2016demonstration}
E.~Mansour, A.~V. Sambra, S.~Hawke, M.~Zereba, S.~Capadisli, A.~Ghanem, A.~Aboulnaga, and T.~Berners-Lee.
\newblock A demonstration of the solid platform for social web applications.
\newblock In \emph{Proceedings of the 25th international conference companion on world wide web}, pages 223--226, 2016.

\bibitem[Mathew et~al.(2021)Mathew, Saha, Yimam, Biemann, Goyal, and Mukherjee]{mathew2021hatexplain}
B.~Mathew, P.~Saha, S.~M. Yimam, C.~Biemann, P.~Goyal, and A.~Mukherjee.
\newblock Hatexplain: A benchmark dataset for explainable hate speech detection.
\newblock \emph{Proceedings of the AAAI Conference on Artificial Intelligence}, 35\penalty0 (17):\penalty0 14867--14875, May 2021.
\newblock \doi{10.1609/aaai.v35i17.17745}.
\newblock URL \url{https://ojs.aaai.org/index.php/AAAI/article/view/17745}.

\bibitem[Mitsikas et~al.(2024)Mitsikas, Sch{\"a}fermeier, and Paschke]{mitsikasmodeling}
T.~Mitsikas, R.~Sch{\"a}fermeier, and A.~Paschke.
\newblock Modeling medical data access with {P}rova.
\newblock \emph{New Frontiers in Artificial Intelligence}, page~35, 2024.

\bibitem[Mozafari et~al.(2020)Mozafari, Farahbakhsh, and Crespi]{mozafari2020bert}
M.~Mozafari, R.~Farahbakhsh, and N.~Crespi.
\newblock Hate speech detection and racial bias mitigation in social media based on bert model.
\newblock \emph{PLOS ONE}, 15\penalty0 (8):\penalty0 1--26, 08 2020.
\newblock \doi{10.1371/journal.pone.0237861}.
\newblock URL \url{https://doi.org/10.1371/journal.pone.0237861}.

\bibitem[Paschke(2011)]{Paschke2011}
A.~Paschke.
\newblock \emph{Rules and Logic Programming for the Web}, pages 326--381.
\newblock Springer, Berlin Heidelberg, 2011.
\newblock ISBN 978-3-642-23032-5.
\newblock \doi{10.1007/978-3-642-23032-5_6}.

\bibitem[Paschke and Bichler(2008)]{PASCHKE2008187}
A.~Paschke and M.~Bichler.
\newblock {Knowledge representation concepts for automated SLA management}.
\newblock \emph{Decision Support Systems}, 46\penalty0 (1):\penalty0 187--205, 2008.
\newblock ISSN 0167-9236.
\newblock \doi{10.1016/j.dss.2008.06.008}.

\bibitem[Paschke and Boley(2014)]{paschkereaction}
A.~Paschke and H.~Boley.
\newblock {Reaction RuleML 1.0 for Distributed Rule-Based Agents in Rule Responder}.
\newblock In \emph{Proceedings of the RuleML 2014 Challenge and the RuleML 2014 Doctoral Consortium, hosted by the 8th International Web Rule Symposium (RuleML 2014)}. CEUR.org, 2014.

\bibitem[Ramachandran et~al.(2020)Ramachandran, Chowdhury, Third, Domingue, Quick, and Bachler]{ramachandran2020towards}
M.~Ramachandran, N.~Chowdhury, A.~Third, J.~Domingue, K.~Quick, and M.~Bachler.
\newblock Towards complete decentralised verification of data with confidentiality: Different ways to connect solid pods and blockchain.
\newblock In \emph{Companion proceedings of the web conference 2020}, pages 645--649, 2020.

\bibitem[Roberts(2017)]{roberts2017content}
S.~T. Roberts.
\newblock \emph{Content Moderation}, pages 1--4.
\newblock Springer International Publishing, Cham, 2017.
\newblock ISBN 978-3-319-32001-4.
\newblock \doi{10.1007/978-3-319-32001-4_44-1}.
\newblock URL \url{https://doi.org/10.1007/978-3-319-32001-4_44-1}.

\bibitem[Sambra et~al.(2016)Sambra, Mansour, Hawke, Zereba, Greco, Ghanem, Zagidulin, Aboulnaga, and Berners-Lee]{sambra2016solid}
A.~V. Sambra, E.~Mansour, S.~Hawke, M.~Zereba, N.~Greco, A.~Ghanem, D.~Zagidulin, A.~Aboulnaga, and T.~Berners-Lee.
\newblock Solid: a platform for decentralized social applications based on linked data.
\newblock \emph{MIT CSAIL \& Qatar Computing Research Institute, Tech. Rep.}, 2016.

\bibitem[Satoh et~al.(2010)Satoh, Asai, Kogawa, Kubota, Nakamura, Nishigai, Shirakawa, and Takano]{satoh2010proleg}
K.~Satoh, K.~Asai, T.~Kogawa, M.~Kubota, M.~Nakamura, Y.~Nishigai, K.~Shirakawa, and C.~Takano.
\newblock {PROLEG}: an implementation of the presupposed ultimate fact theory of {J}apanese civil code by prolog technology.
\newblock In \emph{JSAI international symposium on artificial intelligence}, pages 153--164. Springer, 2010.

\bibitem[Sch{\"a}fermeier et~al.(2022)Sch{\"a}fermeier, Mitsikas, and Paschke]{schafermeier2022modeling}
R.~Sch{\"a}fermeier, T.~Mitsikas, and A.~Paschke.
\newblock Modeling a {GDPR} compliant data wallet application in {P}rova and {AspectOWL}.
\newblock In \emph{RuleML+ RR (Companion)}, 2022.

\bibitem[Wang et~al.(2023)Wang, Hee, Awal, Choo, and Lee]{wang2023evaluating}
H.~Wang, M.~S. Hee, M.~R. Awal, K.~T.~W. Choo, and R.~K.-W. Lee.
\newblock Evaluating {GPT-3} generated explanations for hateful content moderation, 2023.

\end{thebibliography}



\end{document}